\documentclass[amsmath,amssymb,aps,floatfix,prd,a4paper]{revtex4}

\usepackage{amsmath}
\usepackage{graphicx}
\usepackage{float}
\usepackage{subfigure}
\usepackage{wrapfig}
\usepackage{multirow}
\usepackage{tabularx}
\usepackage{array}
\usepackage{url}
\usepackage{amssymb}
\usepackage{systeme}
\usepackage{braket}
\usepackage{caption}
\usepackage{dcolumn}
\usepackage{bm}
\usepackage{epsfig}
\usepackage{amsfonts}
\usepackage{xcolor}

\def\beq{\begin{equation}}
\def\eeq{\end{equation}}
\def\beqa{\begin{eqnarray}}
\def\eeqa{\end{eqnarray}}

\def\gappeq{\mathrel{\rlap {\raise.5ex\hbox{$>$}}
{\lower.5ex\hbox{$\sim$}}}}
\def\lappeq{\mathrel{\rlap{\raise.5ex\hbox{$<$}}
{\lower.5ex\hbox{$\sim$}}}}
\def\Toprel#1\over#2{\mathrel{\mathop{#2}\limits^{#1}}}

\begin{document}

\title{Understanding   the  $K^*/K$ ratio  in  heavy ion collisions}
\author{C. Le Roux,  F. S. Navarra}
\affiliation{$^1$Instituto de F\'{\i}sica, Universidade de S\~{a}o Paulo, 
Rua do Mat\~ao, 1371, CEP 05508-090,  S\~{a}o Paulo, SP, Brazil}
\author{ L. M. Abreu}
\affiliation{ Instituto de F\'isica, Universidade Federal da Bahia,
Campus Universit\'ario de Ondina, 40170-115, Bahia, Brazil}

\begin{abstract}
We study the $K^*$ meson dissociation in heavy ion collisions during the 
hadron gas phase. We use the production and absorption  
cross sections of the $K^*$ and $K$ mesons in a hadron gas, which were 
calculated in a previous work. 
We compute the time evolution of the $K^*$  abundance and the $K^* /K$ ratio 
during the  hadron gas phase of heavy ion collisions.  Assuming a Bjorken type
cooling and using an empirical relation between the freeze-out temperature and the
central multiplicity density, we are able to write $K^* /K$ as a function of 
($ dN /d \eta (\eta =0)$). The obtained function is in very good  agreement with
recent experimental data.

\end{abstract}

\maketitle

\section{Introduction}

In recent heavy ion collision experiments nuclei are accelerated  towards
each other 
with energies of the order of GeV or TeV. These extremely high energies allow 
for the production of a deconfined phase of quarks and gluons. This phase where 
the fundamental particles are able to travel freely is known as the quark gluon 
plasma (QGP) \cite{shu,gm}. It exists for a short time and as the system
expands and cools down, quarks, antiquarks and gluons recombine to form hadrons.
This phase transition back to the hadron phase is also called hadronization.  
The abundances of particles formed during the hadronization depend on the
temperature
and on the  baryon  chemical potential.  After hadronization the system becomes
a hot hadron gas in which inelastic reactions occur,  changing the relative
abundance of the hadrons. The system further expands and cools down until the
point 
when all interactions cease. This is known as kinetic or thermal freeze-out. 
The final yield of hadrons in a collision is influenced not only by their   
production rate at the quark-hadron transition point but also by the
interactions
that they undergo  after hadronization, which might increase or decrease
their abundances. At the thermal freeze-out,  particle abundances are frozen
and the
hadrons flow freely to the detectors.


The $K^*$ meson is a resonance and  may change its abundance also by the strong 
decay $K^* \to K \pi$. This meson has a lifetime of 4 fm/c, smaller than the
duration
of the hadron gas phase, which is believed to be of the order of 10 fm/c.
When the
decay happens in the hadronic medium, the daughter particles ($K$ and $\pi$)
interact
further with other particles in the environment, changing their energy and
momentum,
and even if they can be measured at the end of the heavy ion collision, the
invariant 
mass of the pair is no longer equal to the $K^*$  mass. The $K^*$ which is
no longer
reconstructed is lost and we would observe a reduction in the final yield
of this
resonance, which would then be attributed to the existence of the hadron
gas phase. 
This means that the existence of the hadron gas phase could be tested by
the study of
the abundances of such particles. This idea has been discussed in several
publications
\cite{torra,bleiche,rafeleto,knospe16,stein17}.


From the experimental point of view, the abundance of the $K^*$  meson
can be studied
through the yield  ratio $K^*$/$K$. Experiments have measured it to be
$ 0.33 \pm 0.01$
in Au+Au collisions at $\sqrt{s_{NN}} = 130$ GeV, $0.23 \pm 0.05$ in
Au+Au collisions at  $\sqrt{s_{NN}} = 200$  GeV  at RHIC
\cite{star05,star11}, $ 0.19 \pm 0.05$ at $\sqrt{s_{NN}} = 2.76$ TeV
in Pb+Pb collisions at the 
LHC \cite{alice15,alice17} and very recently \cite{alice20} it
was found to be $0.2 \pm 0.01$ in Pb+Pb at $\sqrt{s_{NN}} = 5.02$ TeV
in collisions at the LHC.
Model calculations suggest that the
lifetime of the hadron gas phase grows with the mass of the colliding
nuclei, with
centrality and with the collision energy. We can see that the ratio $K^* /K$
decreases as the collision energy and/or system size increases, giving support
to the conjecture made
in the previous paragraph. However, in order to reach a firm conclusion a
comprehensive
quantitative calculation must be done. 


In the hadron gas formed in heavy ion collisions, the temperatures range from
$\simeq 175$ MeV, where hadronization takes place, to $\simeq 100$ MeV, where
kinetic freeze-out takes place. The temperature defines the order of magnitude of
the hadron momenta in the gas and also the energy with which hadrons collide in
the medium. The energies of a few hundred MeV's are too high to allow the use
of chiral
perturbation theory and are too low to allow the use of perturbative QCD.
One has to
resort to models involving  mesons and baryons. In principle baryons could be
efficient in absorbing $K^*$'s. Although the coupling constants $B B' K^{(*)}$
(baryon-baryon-strange meson) are relatively small \cite{miru99}, it has been
shown in \cite{torres15} that the interaction cross sections are significant and 
the $K^* N$ total cross section can be as large as 20 mb. On the other
hand, the  
particles which emerge from the hadron gas have low or moderate rapidities and
in this rapidity region there are no remnant baryons from the projectile or from
the target. There are only newly created baryons, which are relatively rare.  
Therefore, here we follow \cite{suhoung} and  neglect $K^*$ interactions
with baryons. 

In Ref. \cite{suhoung}  the authors computed the cross sections of several   
types of interactions suffered by $K^*$ and $K$ mesons in the hadron gas and  
showed that,  due to these interactions and to the strong decay, the final yield
ratio $K^*$/$K$ measured in  central Au+Au collision at
$\sqrt{s_{NN}} = 200$ GeV 
decreases by 37 \%  during the hadron gas phase, resulting in a final
ratio comparable to STAR measurements.  In \cite{suhoung}, the change in the
abundances of the  $K^*$ and $K$ mesons was computed by solving a system of
differential  rate equations which use as input the cross sections for different
interactions involving the $K^*$ and $K$  mesons with each other and with 
the light mesons $\rho$ and $\pi$. The authors found that the leading
processes contributing 
to the abundance dynamics are: $K^* \pi \rightarrow K \rho$,             
$K^* \rho \rightarrow K \pi$ and $K^* \rightarrow K \pi$, as well as
the inverse ones. 

\begin{figure}[!ht]
\begin{tabular}{ccc}
  \includegraphics[width=.50\linewidth]{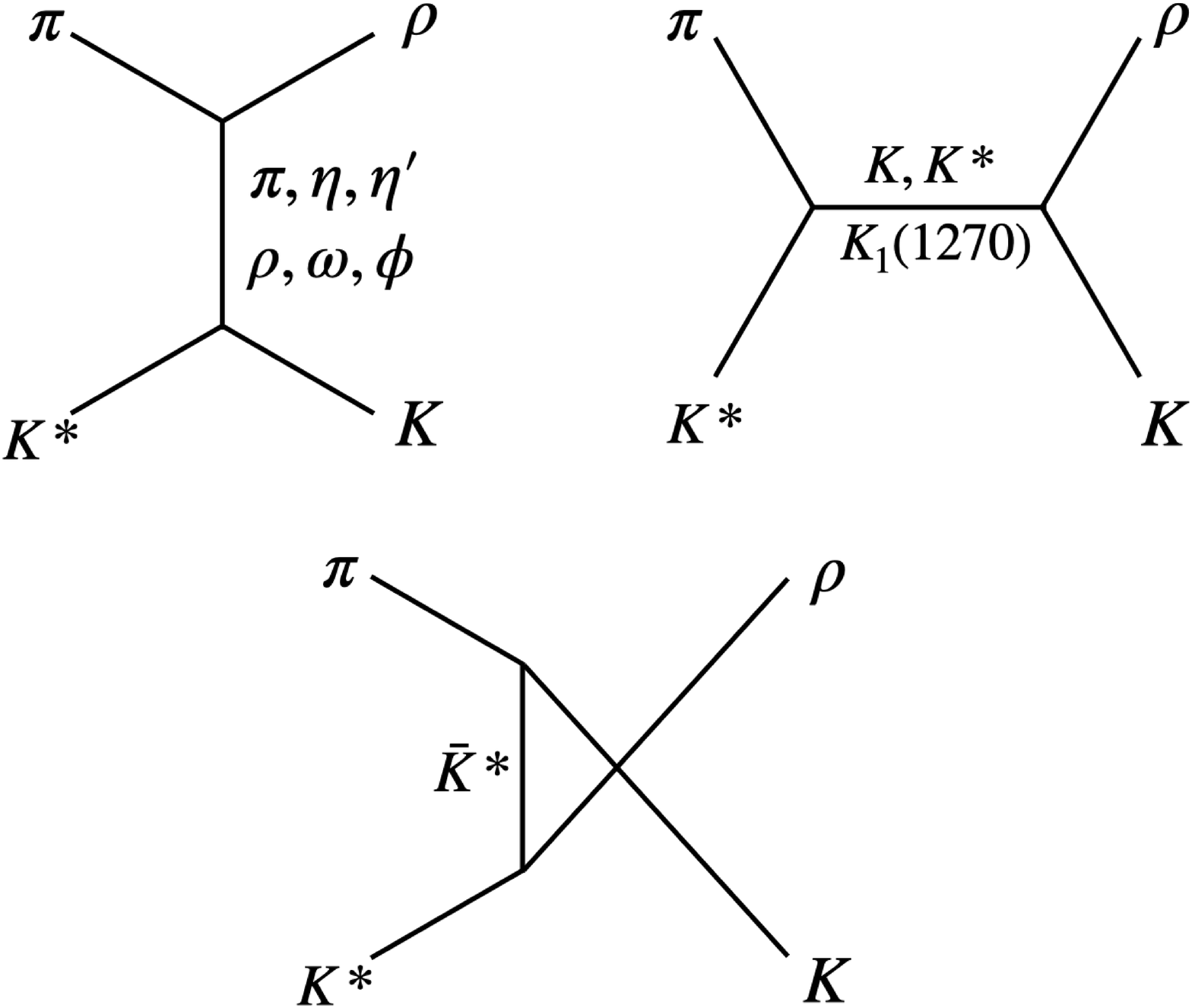}& \,\,\, &
  \includegraphics[width=.50\linewidth]{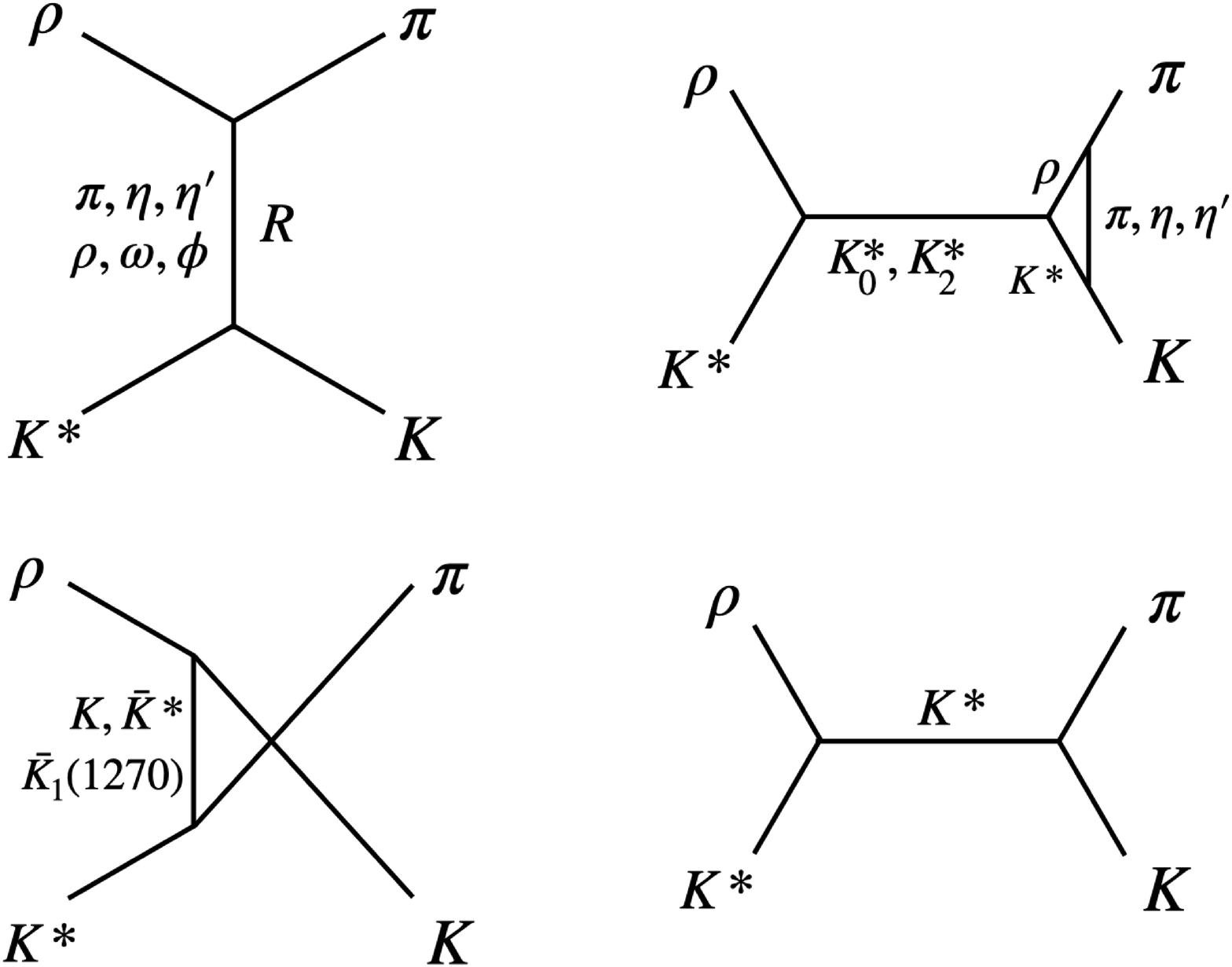} \\
  (a) & \,\,\, & (b)
   \end{tabular}
\caption{Diagrams for the relevant processes considered in the calculation
of the cross sections in Ref. \cite{abreu}. a) $K^* \pi \to \rho K$
reactions. $R$ represents the resonances $h_1(1170)$, $h_1(1380)$, $f_1(1285)$,
$a_1(1260)$ and $b_1(1235)$.  b) $K^* \rho \to K \pi$ reactions.}
\label{diagramas}
\end{figure}

In \cite{suhoung} some interaction mechanisms that might be relevant
were not included in the calculations.  In a subsequent work \cite{abreu}
the cross sections 
for production and annihilation of $K^*$ and $K$ mesons were
recalculated with the inclusion of new reaction mechanisms. The relevant 
Feynman diagrams for the  $K^* \pi \to \rho K$  and $K^* \rho \to K \pi$
reactions are shown in Fig. \ref{diagramas}.

The most important  (but not the only ones)  changes made in \cite{abreu} are:
\vskip5mm
\noindent
I) Inclusion of anomalous parity vector-vector-pseudoscalar (VVP)
  interactions. 
\vskip5mm
\noindent
II) Inclusion of the exchange of axial resonances $K_1(1270)$,
  $h_1(1170)$,
$h_1(1380)$, $f_1(1285)$, $a_1(1260)$ and $b_1(1235)$ in the s and t channels. 

\vskip5mm
Modification I) introduces new vertices, modifies several Feynman
diagrams and changes the amplitudes of
all processes discussed previously in \cite{suhoung}.  In
Refs. \cite{oh,babi05,torres14},
it was shown that interaction terms with anomalous parity couplings 
have a strong impact on the corresponding cross sections.
The relevance of such anomalous terms in the
determination of the abundance of $X(3872)$ in heavy ion
collisions was computed in Ref. \cite{abreu16}. In \cite{abreu} these  
interaction terms were found to be relevant also in the calculation of
$K^*$ absorption processes.
Modification II) introduces several new diagrams. 
The presence of the resonance $K_1(1270)$, for example,  had been found
to be important \cite{geng07} in describing the invariant mass
distribution of the process  $K^- p \to K^- \pi^+ \pi^- p$ at
$\sqrt{s_{NN}} = 63 $ GeV
measured by the WA3 collaboration at CERN \cite{daum81}. In \cite{abreu} it was
seen that the diagram with  $K_1(1270)$ in the s-channel is the most important
contribution to the absorption process $\pi K^*  \to \omega K$ and also to
the production process $\rho K \to \pi K^*$.

The results in Ref. \cite{abreu} show that the new mechanisms are rather         
significant, changing the cross sections up to one or two orders of
magnitude in      
some cases, suggesting that these new cross sections would result in a
very different 
dynamics for the abundances of $K^*$  and $K$ mesons. A comparison
between the results obtained in Refs. \cite{suhoung} and  in \cite{abreu} is
presented in Fig.  \ref{fit}, where we show the thermally averaged cross
sections  (see below) of the main processes of absorption and regeneration of
$K^*$.  From the figures we see that the cross sections found in \cite{abreu}
are much larger than those found in \cite{suhoung}, both for absorption and for
regeneration of $K^*$. 

In this work  we use the improved cross sections of \cite{abreu}  and solve  
the differential rate equations proposed in Ref. \cite{suhoung}, obtaining  
the $K^*/K$ ratio as a function of the proper time.  We use a Bjorken type
cooling to connect the proper time and the temperature. The evolution stops at
the freeze-out temperature $T_f$. Finally, we use the empirical relation between
$T_f$ and the central multiplicity density found in 
\cite{alice13}, to obtain a direct relation between the $K^*/K$  ratio and
$ dN / d\eta (\eta = 0)$. The obtained relation is in very good agreement with
experimental data. In the next section we briefly describe the formalism and in
the following section we present our results and compare them with experimental
data. 

\section{Formalism}

\subsection{Thermal cross sections}

In \cite{suhoung} the interactions of $K$ and $K^*$ with light non-strange mesons
were described by effective Lagrangians of the type $\mathcal{L}_{PPV}$ and
$\mathcal{L}_{VVV}$, where $P$ and $V$ are pseudoscalar and  vector mesons,
respectively.
The Lagrangians were obtained from free pseudoscalar and vector meson
Lagrangians by
introducing the minimal substitution.
In \cite{abreu}, in addition to these Lagrangians,
the  Lagrangian $\mathcal{L}_{VVP}$
was included, representing the so called ``anomalous parity'' interactions.
From the
Lagrangians it is straightforward to evaluate the amplitudes of $K^*$
absorption by
pions, kaons, $\rho$'s and by $K^*$'s.  In order to take the finite size of
the hadrons
into consideration when evaluating amplitudes, one uses  form factors at each
interaction
vertex. These form factors contain a cut-off parameter. In \cite{suhoung} the
authors took 
a value taken from previous phenomenological analises \cite{brown91}.  With the 
amplitudes
it is easy to compute the cross sections of the corresponding processes.  With 
the same
Lagrangians one can calculate all the interaction cross sections of kaons.  
Moreover, with the
use of detailed balance one can calculate the inverse processes, i.e. one   
can compute the $K^* + \pi \to K + \rho$ and $ K + \rho \to K^* + \pi$ cross
sections. Finally, one has to consider the processes
$K + \pi \to K^*$ and $ K^* \to K + \pi$. 
The authors of \cite{suhoung} also found that the cross section for the    
formation of the $K^*$ meson from pions and K mesons  is not small at all,
compared to cross sections for other processes.

All the reactions mentioned above happen within a hadron gas at temperatures
ranging from 100 to 200 MeV.
These temperatures determine the collision energies. Moreover the densities
of the colliding particles
are determined by the temperature. Therefore, in this context, the most relevant
dynamical quantity is
the thermally averaged cross section. For a process $a+b \to c+d$ it is
defined as:
\begin{equation}
  \braket{\sigma_{ab \rightarrow cd} v_{ab}} = \frac{1}{1+\delta_{ab}}	
  \frac{\int d^3 \vec{p_a} d^3 \vec{p_b} f_a(\vec{p_a}) f_b(\vec{p_b})	
    \sigma_{ab \rightarrow cd} v_{ab}}{\int d^3 \vec{p_a} d^3 \vec{p_b}
    f_a(\vec{p_a}) f_b(\vec{p_b})},
  \label{tax}
\end{equation}
where $v_{ab}$ is the relative velocity between the initial particles
$$
v_{ab} =\sqrt{(p_a \cdot p_b)^2-m^2_am^2_b}/(E_a E_b)
$$
and $f_i(\vec{p_i}) $ is the thermal momentum distribution of particle $i$,
which is given by 
a Bose-Einstein  distribution:
$$
f_i(\vec{p_i}) = \frac{1}{ e^{\sqrt{\vec{p_i}^2+m_i^2}/T} -1}.
$$
The production and absorption rates of $K^*$ or $K$ obviously depend on the
densities of particles in the hadron gas at proper time $\tau$ which are given by 
\begin{equation}
\label{densidades}
n_i(\tau) = \frac{g_i}{2 \pi^2} \, \int_0^{\infty}
\frac{ p^2 dp}
     { e^{\sqrt{p^2_i + m^2_i}/T(\tau)}  -  1    }
       \simeq 
\frac{g_i}{2 \pi^2} m_i^2 T(\tau) K_2 \left( \frac{m_i}{T(\tau)} \right), 
\end{equation}
where $g_i$ is the degeneracy factor of meson $i$ and $m_i$ its mass. 
$K_2(\tau)$ is the modified Bessel function of the second kind and
$T(\tau)$ is the temperature.  The total number of particles of 
species $i$, $N_i(\tau)$,  is obtained by
multiplying the density given by (\ref{densidades}) by the system   
volume $V(\tau)$. At last, the thermally averaged decay width of $K^*$ 
was computed using the following expression
introduced in Ref. \cite{suhoung}:
\begin{equation}
\label{decay}
        \braket{\Gamma_{K^*}}=\Gamma_{K^*}(m_{K^*}) \frac{K_1
          \left( \frac{m_{K^*}}{T(\tau)}
          \right )}{ K_2 \left( \
\frac{m_{K^*}}{T(\tau)} \right)},
\end{equation}
where $K_1$ and $K_2$ are the modified Bessel functions of the first and
second kind, $T(\tau)$  is the temperature as a function of proper time
$\tau$, $m_{K^*}$
is the mass of $K^*$ and $\Gamma_{K^*}$, its decay width, which was
computed as:
\begin{equation}
  \Gamma_{K^*} (\sqrt{s}) = \frac{ g^2_{\pi K K^{*}} }{2 \pi s} p_{cm}^3
  (\sqrt{s}),
\end{equation}
with $g_{\pi K K^{*}}$ being the coupling constant, $p_{cm}$  the momentum
at the center of mass frame and $s$ the Mandelstam variable. 

\begin{figure}[!ht]
\begin{tabular}{cc}
  \includegraphics[width=.45\linewidth]{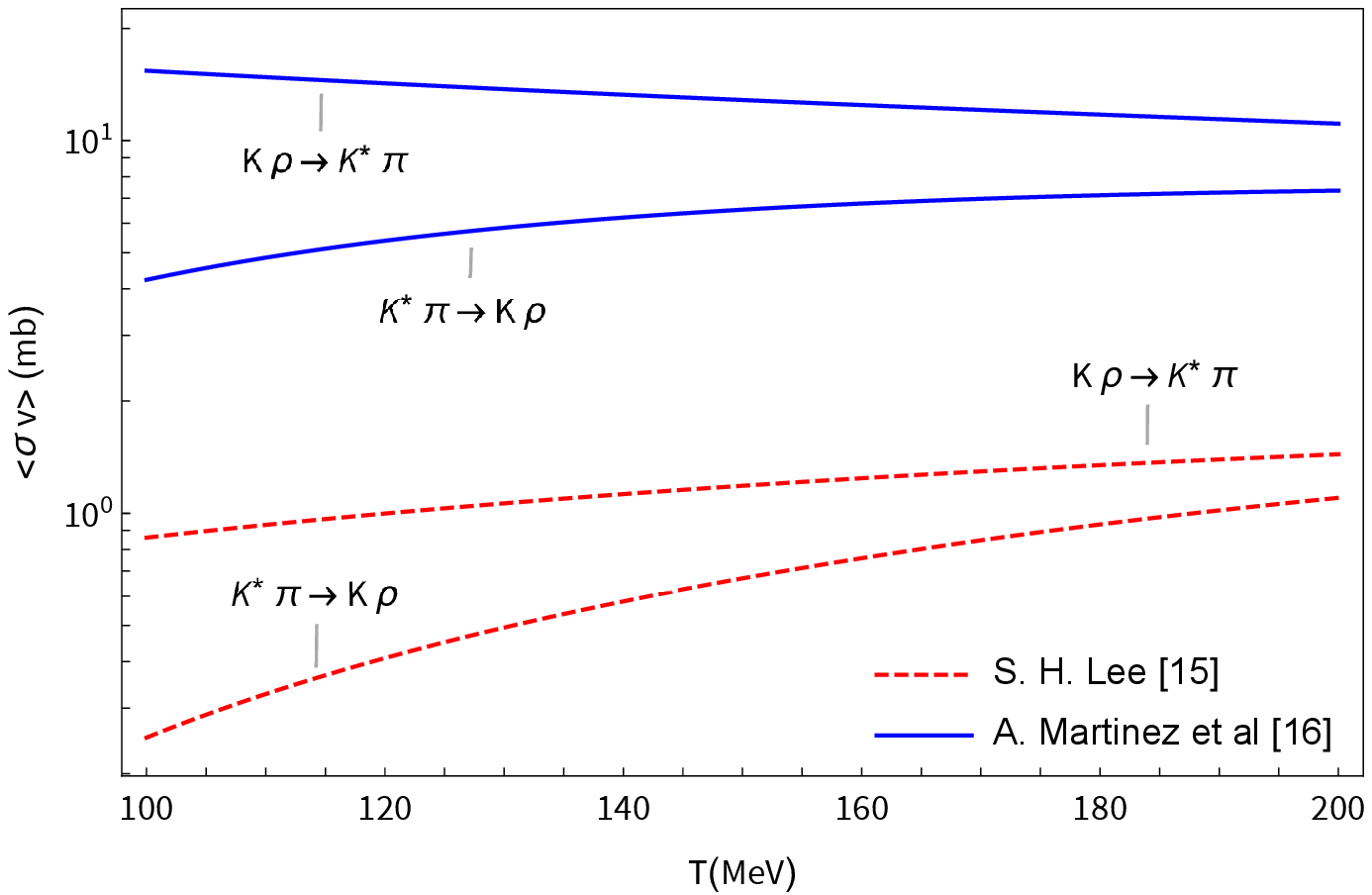}&
  \includegraphics[width=.45\linewidth]{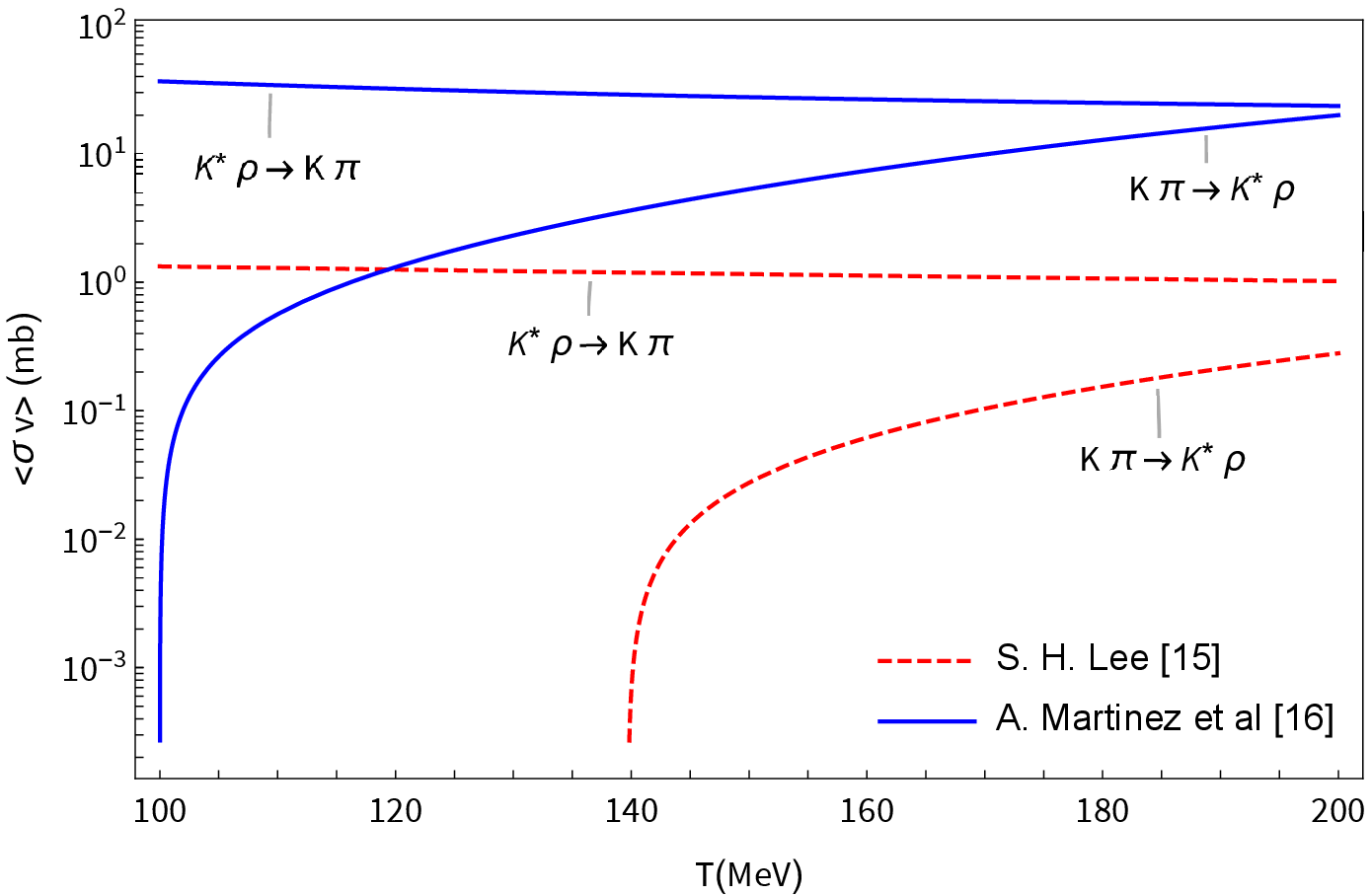}\\ 
  (a) & (b) 
   \end{tabular}
\caption{Comparison between the cross sections obtained in Ref. \cite{suhoung}
  and those obtained in Ref. \cite{abreu}. The lines are obtained with
  Eq. (\ref{param}) which is a parametrization of the results obtained in the
  mentioned papers.
  a) $K^* \pi \to \rho K$ 
reactions. b) $K^* \rho \to K \pi$ reactions.}
\label{fit}  
\end{figure}

For our purposes it is not necessary to recalculate all the thermally
averaged cross
sections, $\braket{\sigma_{ab \rightarrow cd} v_{ab}}$, which are smooth
functions
of the temperature. It is enough to parametrize the results obtained in Refs.
\cite{suhoung} and \cite{abreu} by the polinomial functions which are shown
in the Appendix. The resulting parametrizations are shown  in  Fig. \ref{fit}.

In Fig. \ref{fit} we can compare the results obtained in \cite{suhoung}
with those obtained in \cite{abreu}. The inclusion of modifications I and II
increased the cross sections tipically by one order of magnitude. In both
approaches the absorption of $K^*$ is stronger than its production. However, 
with the formalism considered  in \cite{abreu}, at higher temperatures we
observe the dominance of the processes of creation of $K^*$. So, when the
hadron gas starts its expansion at high temperatures, we expect to see first
the growth of the $K^*$ multiplicity which is later followed by its reduction.
In contrast, with the formalism of \cite{suhoung} we only see a monotonic
reduction of the $K^*$ multiplicity.

\subsection{Evolution equations} 

With the ingredients presented in the previous subsection, it is possible to
write
rate equations, which describe the time evolution of the $K^*$ and $K$
multiplicities,
incorporating the gain and loss terms due to production and absorption
respectively. These equations are: 
\begin{align} 
	\frac{dN_{K^*}}{d \tau} =& \braket{\sigma_{K \rho \rightarrow K^* \pi} v_{K \rho}}n_{\rho}(\tau)N_K(\tau)-\braket{\sigma_{K^* \pi \rightarrow K \rho} v_{K^* \pi}} n_\pi (\tau) N_{K^*} (\tau) + \braket{\sigma_{K \pi \rightarrow K^* \rho} v_{K \pi}}n_{\pi}(\tau)N_K(\tau) \notag
\\ &- \braket{\sigma_{K^* \rho \rightarrow K \pi} v_{K^* \rho}}n_{\rho}(\tau)N_{K^*}(\tau) + \braket{\sigma_{\pi \rho \rightarrow K^* \bar{K}} v_{\pi \rho}}n_{\pi}(\tau)N_\rho(\tau) - \braket{\sigma_{K^* \bar{K} \rightarrow \rho \pi} v_{K^* \bar{K}}}n_{\bar{K}}(\tau)N_{K^*}(\tau) \notag 
\\ &+ \braket{\sigma_{\pi \pi \rightarrow K^* \bar{K}^*} v_{\pi \pi}}n_{\pi}(\tau)N_\pi(\tau)-\braket{\sigma_{K^* \bar{K}^* \rightarrow \pi \pi} v_{K^* \bar{K}^*}}n_{\bar{K}^*}(\tau)N_{K^*}(\tau) + \braket{\sigma_{\rho \rho \rightarrow K^* \bar{K}^*} v_{\rho \rho}}n_{\rho}(\tau)N_\rho(\tau) \notag
	\\ &- \braket{\sigma_{K^* \bar{K}^* \rightarrow \rho \rho} v_{K^* \bar{K}^*}}n_{\bar{K}^*}(\tau)N_{K^*}(\tau) + \braket{\sigma_{K \pi \rightarrow K^*} v_{K \pi}}n_{\pi}(\tau)N_K(\tau) - \braket{\Gamma_{K^*}} N_{K^*}(\tau), \notag \\	
	\frac{dN_{K}}{d \tau} =& \braket{\sigma_{\pi \pi \rightarrow K \bar{K}} v_{\pi \pi}}n_{\pi}(\tau)N_\pi(\tau)-\braket{\sigma_{K \bar{K} \rightarrow \pi \pi} v_{K \bar{K}}}n_{\bar{K}}(\tau)N_{K}(\tau)+\braket{\sigma_{\rho \rho \rightarrow K \bar{K}} v_{\rho \rho}}n_{\rho}(\tau)N_\rho(\tau) \notag \\
	&-\braket{\sigma_{K \bar{K} \rightarrow \rho \rho} v_{K \bar{K}}}n_{\bar{K}}(\tau)N_{K}(\tau)+\braket{\sigma_{K^* \pi \rightarrow K \rho} v_{K^* \pi}} n_\pi (\tau) N_{K^*} (\tau) - \braket{\sigma_{K \rho \rightarrow K^* \pi} v_{K \rho}}n_{\rho}(\tau)N_K(\tau) \notag \\
	&+ \braket{\sigma_{K^* \rho \rightarrow K \pi} v_{K^* \rho}}n_{\rho}(\tau)N_{K^*}(\tau)-\braket{\sigma_{K \pi \rightarrow K^* \rho} v_{K \pi}}n_{\pi}(\tau)N_K(\tau) +\braket{\sigma_{\pi \rho \rightarrow K^* \bar{K}} v_{\pi \rho}}n_{\pi}(\tau)N_\rho(\tau) \notag \\
	&-\braket{\sigma_{K^* \bar{K} \rightarrow \rho \pi} v_{K^* \bar{K}}}n_{\bar{K}}(\tau)N_{K^*}(\tau)+\braket{\Gamma_{K^*}} N_{K^*}(\tau)-\braket{\sigma_{K \pi \rightarrow K^*} v_{K \pi}}n_{\pi}(\tau)N_K(\tau) .
	\label{sistemao}
\end{align}
 
The above  equations include all relevant creation and annihilation reactions. However,
as showed in Refs. \cite{suhoung} and \cite{abreu}, some of them have very small
thermally averaged cross sections and can be safely neglected. The really important   
interactions of the $K^*$ meson according to both references are the following:
\begin{align}
  \label{processes}
	K^* \rho &\rightarrow K \pi,  \notag \\
	K^* \pi &\rightarrow K \rho,  \notag \\
	K^* &\rightarrow K \pi, 
\end{align}
as well as the respective inverse processes. This should not be surprising
since
$\pi$'s are the most abundant particles in a hadron gas and $\rho$'s are vector
particles and, as discussed above, have a large interaction cross section
with  other vector particles. Restricting ourselves to the processes above,
the system of differential equations Eq.(\ref{sistemao})  can be written as:
\begin{align}
  \label{sistema}
  \frac{dN_{K^*}(\tau)}{d\tau}&=\gamma_K N_K (\tau)
  - \gamma_{K^*} N_{K^*}(\tau), \nonumber\\
    \frac{dN_{K}(\tau)}{d\tau}&=-\gamma_K N_K (\tau) + \gamma_{K^*} N_{K^*}(\tau), 
\end{align}
where $N_K$ and $N_{K^*}$ are the abundances of K and $K^*$ mesons respectively.
They are functions of the proper time $\tau$. The factors $\gamma_K$ and
$\gamma_{K^*}$ depend on the interaction cross sections and the light meson
densities in the following way:
\begin{align} 
  \gamma_{K}&=\braket{\sigma_{K\pi\xrightarrow{}K^*\rho}v_{K\pi}}n_{\pi} +
  \braket{\sigma_{K\rho\xrightarrow{}K^*\pi}v_{K\rho}}n_{\rho}+
  \braket{\sigma_{K \pi \xrightarrow{} K^* }v_{K \pi}} n_\pi,  \nonumber\\
  \gamma_{K^*}&=\braket{\sigma_{K^*\rho\xrightarrow{}K\pi}v_{K^*\rho}}n_{\rho}+
  \braket{\sigma_{K^*\pi\xrightarrow{}K\rho}v_{K^*\pi}}n_{\pi}+
  \braket{\Gamma_{K^*}}.
\label{gammas}
\end{align}
It is interesting to consider the limiting case where the temperature and
light meson densities stay constant in time.  
In this case  $\gamma_{K^*}$  and  $\gamma_K$ are constant and the
system (\ref{sistema}) can be solved analytically giving
the following result:
\begin{align}
  \label{anal}
  N_{K^*}(\tau)&=\frac{\gamma_K}{\gamma}N^0 +
  \left (N_{K^*}^0- \frac{\gamma_K}{\gamma}
   N^0 \right )e^{-\gamma(\tau - \tau_h)},  \nonumber \\
   N_{K}(\tau)&=\frac{\gamma_{K^*}}{\gamma}N^0+ \left
   (N_{K}^0- \frac{\gamma_{K^*}}{\gamma} N^0 \right )
   e^{-\gamma(\tau - \tau_h)},
\end{align}
where $N^0 = N_{K^*}^0+N_K^0$, i.e., the sum of the initial abundances of 
$K$ and $K^*$.   Moreover, $\gamma=\gamma_{K^*}+\gamma_K$, as computed in    
expressions (\ref{gammas}). At the hadronization time, $\tau_h$, 
the system of $K^*$'s and $K$'s starts to evolve and collide with the light
particles from the reservoir which is kept at constant temperature. At large
times the $N_{K^*}^0$ and $N_K^0$ reach their asymptotic constant values. This
is our operational definition of chemical equilibrium. 

Once we define the temperature evolution (``cooling'') of the hadron gas
$T(\tau)$ and the initial conditions $N_{K^*}(\tau_h)$ and  $N_K(\tau_h)$,
the system of differential equations (\ref{sistema}) can be solved,
yielding $N_{K^*}$,  $N_K$ and the ratio $R(\tau)$ :
\begin{equation}
  \label{ratio}
        R(\tau)=\frac{N_{K^*}}{N_{K}} = \frac{K^*}{K} . 
\end{equation}
We  follow the time evolution of the abundances until the kinetic
freeze-out of the gas, which is defined by the freeze-out temperature $T_f$ 
and occurs at time $\tau_f$. Assuming that the hadronic system undergoes a
Bjorken-like expansion, we may write:
\begin{equation}
T = T_h \left(\frac{\tau_h}{\tau}\right)^{1/3}, 
\label{bjor}
\end{equation}
where $T_h=175$ MeV is the universal hadronization temperature discussed
above and $\tau_h$  is the hadronization time, which may change
from system to system.  We take the  above expression at the particular
freeze out time, $\tau_f$ and freeze-out temperature, $T_f$, and invert it
to obtain:
\begin{equation} 
\tau_f = \tau_h  \left( \frac{T_h}{T_f} \right)^3.
\label{bjorf}
\end{equation}
We  solve (\ref{sistema}) until $\tau_f$ and compute the ratio
$R[\tau_f(T_f)]$.
As it was pointed out long ago \cite{hama92}, the kinetic freeze-out temperature
is not an universal constant. It depends on the size of the hadronic system and
hence on  the collision energy, on the mass number of the colliding nuclei
and on the centrality of the collision.  A recent blastwave fit analysis made
by the ALICE Collaboration \cite{alice13} has confirmed that the  kinetic
freeze-out temperature decreases with the system size, customarily associated to 
the multiplicity density of charged particles, $d N / d \eta $,  measured at  
midrapidity. The empirical relation between $T_f$ and $\mathcal{N} $ found in
\cite{alice13} can be parametrized as: 
\begin{equation}
T_f  = \frac{T_{f0}}{\mathcal{N}^{a}}, 
\label{cris}
\end{equation}
where $T_{f0}$ and $a$ are constants. Inserting (\ref{cris}) into (\ref{bjorf})
we find that
\beq
\tau_f \propto \mathcal{N}^{3 a}.
\label{relf}
\eeq
This relation
tells us that $\mathcal{N}$ gives a measure of the duration of the
hadronic phase. Larger systems (with larger $\mathcal{N}$) live longer. 
Using the obtained $\tau_f$ to determine the end of the evolution of
(\ref{sistema})
we find $R$ as a function  $\mathcal{N}$. The function $R$ can then be  
directly compared  with the data on $R$ versus $\mathcal{N}$  presented
very recently in \cite{alice20}. This will be done in the next section.

\section{Results and Discussion}  

From what was said above we see that the final multiplicities of $K^*$ and
$K$ may depend on: i) the collision dynamics, i.e., on the production and
absorption cross sections discussed above; ii) the initial conditions of the
evolution equations (\ref{sistema}), i.e., the initial values of $N_{K^*}$
and $N_K$; iii) the expansion dynamics, i.e., the cooling function $T(\tau)$
and iv) the system size, characterized by $d N / d \eta (\eta=0)$.

\begin{figure}[!ht]
\vskip5mm
\begin{tabular}{ccc}
 \includegraphics[width=.50\linewidth]{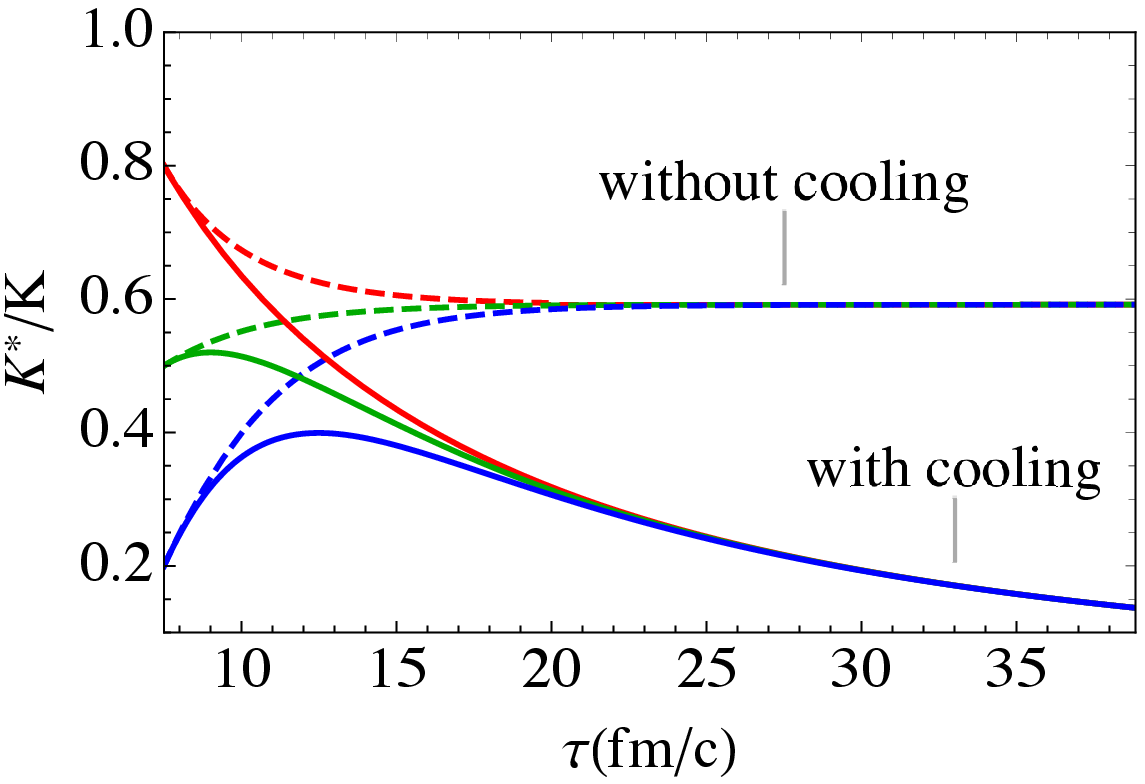}& \,\,\, &
  \includegraphics[width=.50\linewidth]{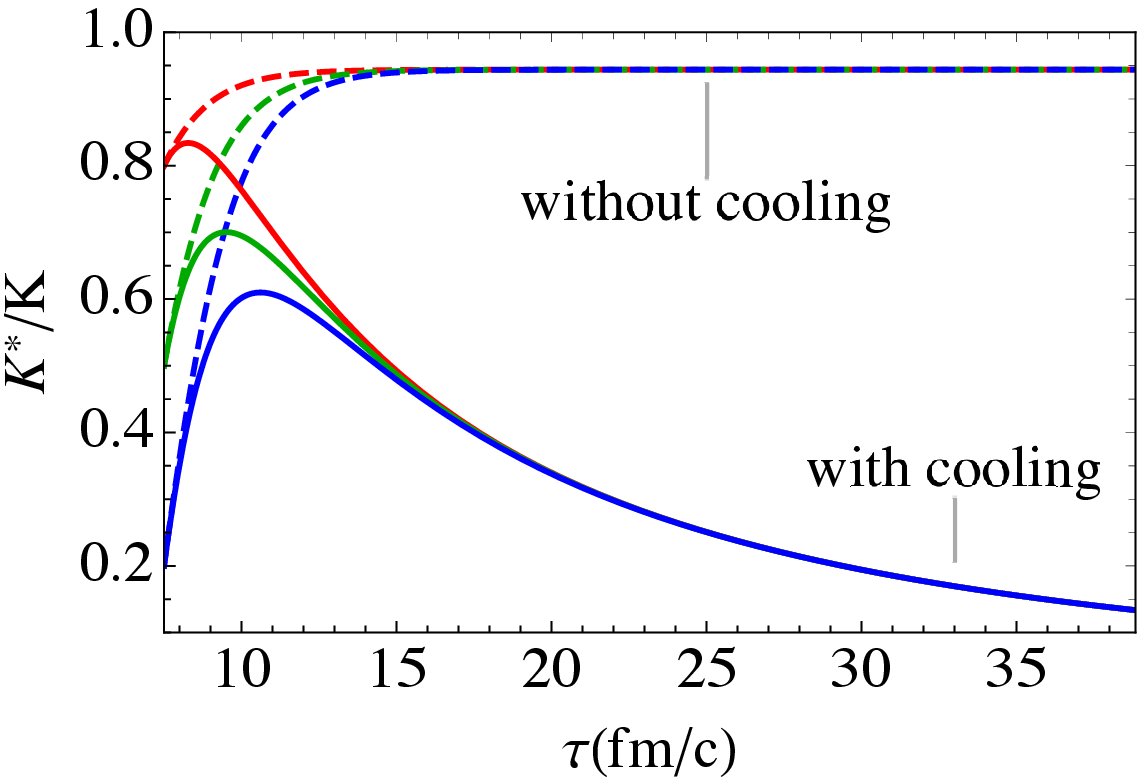} \\
  (a) & \,\,\, & (b)
   \end{tabular}
\caption{$K^* / K $ ratio as a function of the proper time $\tau$. Dashed lines 
correspond to the initial conditions 0.2, 0.5 and 0.8 and no cooling.  
Solid lines correspond to the initial conditions 0.2, 0.5 and 0.8 and
cooling.   a) Cross sections from  S. H. Lee \textit{et al} \cite{suhoung}.
b) Cross sections from  A. Martinez \textit{et al}  \cite{abreu}. 
}
\label{figratio}
\end{figure}

We  solve the equations (\ref{sistema}) using as input the cross
sections calculated in \cite{suhoung} and in \cite{abreu}. The initial temperature
is $T_h =175$ MeV and the initial conditions are $K^* / K  = $  $0.2$, $0.5$ and
$0.8$. The results are shown in Fig.  \ref{figratio}. On the left (right) panel
the inputs are from Ref. \cite{suhoung} (\cite{abreu}). 

First we observe that, as anticipated from Eqs. (\ref{anal}), when there is no
cooling the system evolves to an asymptotic state where the abundances become
constant. When cooling (\ref{bjor}) is included, the ratio $K^* / K $ drops and  
at typical freeze-out times of 20 - 25 fm/c reaches 0.2 - 0.3.  These numbers are
close to the measured ones. This suggests that a  cooling  faster
than (\ref{bjor}),
such as the Hubble-like cooling discussed in \cite{singh20,ghosh20}, is probably
incompatible with data. Another interesting aspect of the figure is that,
even with
cooling, after some time of  evolution the $K^* / K $  ratio becomes the same for
all initial conditions. Comparing the left and right panels we observe the effect
of changing the microscopic cross sections from those calculated in
\cite{suhoung}
to those calculated  in \cite{abreu}.  When there is no cooling the ratio shown
on the left (with the inputs from \cite{suhoung}) is significantly smaller than
the one on the right (with the inputs from \cite{abreu}). This is a
consequence of
Fig. \ref{fit}: at higher temperatures, with \cite{abreu} the cross section for 
$K^*$ production is bigger and so is the ratio $R$. It is also for this reason
that
on the right panel we observe a growth, in some cases very pronounced, of all
lines at early times. 

In \cite{abreu}  all the
cross sections are bigger, all the reactions happen faster and hence the system
looses sooner the memory of the initial conditions (the three lines become a
single
line). Interestingly, at very long times in both cases (right and left panels)
the ratio goes to the same value.

From  Fig. \ref{figratio}  
it is clear  that the new reactions mechanisms considered in \cite{abreu} have 
an impact on the evolution of the abundances of $K^*$ and K mesons in the
hadronic medium. They predict a time evolution of the abundances which is
considerably different from previously thought: there
is an initial increase in the yield ratio which would not exist without
taking into account all the possible mechanisms for the processes in
(\ref{processes}). Unfortunately, the differences with respect to the previous
calculations of Ref. \cite{suhoung} are washed out during the evolution and in
the end the improved cross sections lead to a final yield ratio very close to
that computed in Ref. \cite{suhoung}.

In order to understand this behavior, it is important to notice from Fig. 
\ref{fit} that even though the cross sections for the  
annihilation of $K^*$ are larger in Ref. \cite{abreu} than in Ref.            
\cite{suhoung}, those for the creation of $K^*$ are also larger. For example,  
Fig. \ref{fit}b  clearly shows that in the case of the creation of
$K^*$ through $K \pi \rightarrow K^* \rho$, the cross sections from 
Ref. \cite{abreu} are one order of magnitude larger than those from 
\cite{suhoung} and, as time passes, i.e., the gas cools down, the     
difference between them decreases considerably. The opposite goes for
the creation of $K^*$ through   
$K \rho \rightarrow K^* \pi$ (Fig.  \ref{fit}a) but the 
difference  between the cross section in Ref. \cite{abreu} and
Ref. \cite{suhoung} in this case is much smaller.

\begin{figure}[!ht]
\vskip5mm
\begin{center}
      \includegraphics[width=0.5\textwidth]{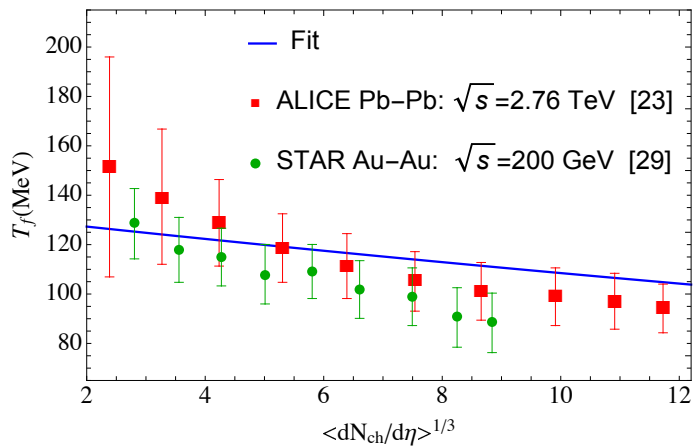}
\end{center}
\caption{Freeze-out temperature as a function of
  $\left[ d N / d \eta (\eta=0)\right]^{1/3}$. The circles are the result
  of the blastwave fits of data on Pb + Pb collisions at
  $\sqrt{s_{NN}} = 2.76$ TeV taken by the ALICE Collaboration \cite{alice13}.
  The squares represent blastwave fits of data on Au + Au collisions
  $\sqrt{s_{NN}} = 200$ GeV taken by the STAR Collaboration \cite{star05}.
The line represents the expression (\ref{chiafit}). }
   \label{freet}
\end{figure}

\begin{figure}[!ht]
\vskip5mm
\begin{center}
      \includegraphics[width=0.60\textwidth]{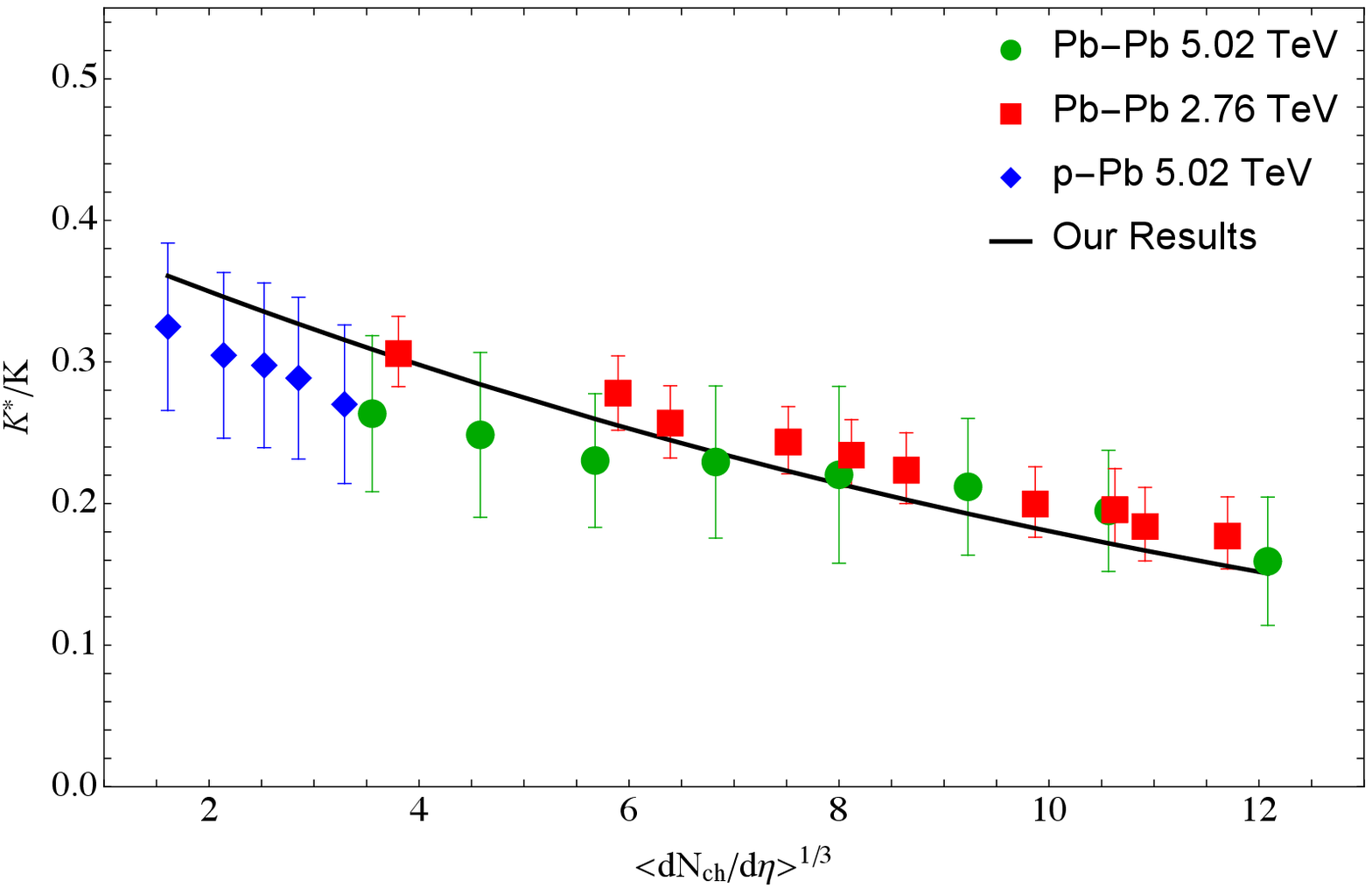}
\end{center}
\caption{$K^* / K$ as a function of
  $\left[ d N / d \eta (\eta=0)\right]^{1/3}$. Data are from \cite{alice20}. 
}
\label{radata}
\end{figure}

In order to compare our results with data, we will make use of the connection
established in \cite{alice13} between $T_f$ and $\mathcal{N}$. Although the
power law fit (\ref{cris}) is very useful because it leads immediately to
(\ref{relf}), a somewhat better fit of the points shown in \cite{alice13} can
be obtained with the form: 
\begin{equation}
T_f  = {T_{f0}} \, e^{- b \, \mathcal{N}},
\label{chiafit}
\end{equation}
where $T_{f0} = 132.5$ MeV and $ b = 0.02$.
The above expression is compared to the data points
from \cite{alice13} in Fig. \ref{freet}. We emphasize that Eq. (\ref{chiafit})
is not the result of a global best $\chi^2$ fit. We try to get a better
description of the higher energy data points, which will be relevant for the study
of the $K^* /K$ ratio measured at the LHC. The STAR points are shown just for
comparison.  
We first choose the
system under consideration, fixing $\mathcal{N}$. This determines the
freeze-out temperature, $T_f$,  and the  endpoint of the evolution, $\tau_f$.
Then, we read the ratio $K^* / K$ from Fig. \ref{figratio}.  Finally, we
plot $K^* / K$ as a function of $\mathcal{N}$ and compare the results with
the data compilation published in \cite{alice20}.  The comparison is presented in
Fig. \ref{radata}. 

As it can be
seen in Fig. \ref{figratio}, the longer the hadronic system lasts, the
smaller is the
ratio $R$. Indeed, for each (increasing) value of $\mathcal{N}$ we
stop the evolution
at an (increasing) value of $\tau$ (which is $\tau_f$) and read from
Fig. \ref{figratio} a (decreasing) value of the ratio $K^* / K$.

There is a strong correlation between Fig. \ref{freet} and Fig. \ref{radata}. 
A steeper function in the first figure implies a steeper function in the second. 
In fact $R \simeq  T_f$.   
Interestingly, the data seem to exclude a flat horizontal line in Fig.
\ref{freet}, i.e., a freeze-out temperature which is universal, independent
of the system size.

Knowing that the existence of a hadron gas phase leads to a reduction in the
ratio $R = K^* / K$, the systematic study of this ratio in different collisions
and at different energies will help us in better determining the properties
of the hadron gas. In proton-proton collisions, where there is no hadron
gas and hence no $K^*$ absorption, $R$ should be maximal. Moving to p-A and
A-A collisions we expect to see the formation of a larger and
longer-living hadron gas. Also, when we move to larger systems we observe a          
growth of the multiplicity of produced particles and also of the
multiplicity density in the central rapidity region                         
$\mathcal{N} =  d N / d \eta (\eta=0)$, which is usually taken as a measure
of the size of the system.
 
In our approach to study the ratio $R = K^* / K$ we made some simplifications.
The goal was to determine which ingredients are  really crucial to understand
the observed behavior. One of the simplifications was to neglect the volume of
the system. The colliding systems mentioned in  Fig. \ref{radata} are different
and so are the corresponding hadronic gases, which have different volumes. In
our study these differences are partly considered in Eq. (\ref{chiafit}).    
Moreover the details of the light flavor composition of these different systems
were not taken into account. In each of the systems considered in Fig. \ref{radata}
the $\pi$ and $\rho$ finally measured multiplicities are different. In thermal
models, this difference is usually accounted by the fugacity factor, $\gamma$,
which should appear multiplying the right side of Eq. (\ref{densidades}).
We have taken
$\gamma=1$ for $\pi$'s and $\rho$'s in $p-Pb$ and $Pb-Pb$ collisions. In previous
studies with thermal models it was shown that, in $Pb-Pb$ collisions, we could
have $\gamma_{\pi} \simeq 1.3$ and $\gamma_{\rho} \simeq 1.2$. Changes in these
quantities would bring changes in Eqs. (\ref{gammas}) and (\ref{sistema}). We have
checked that, using these values for  $\gamma_{\pi}$ and $\gamma_{\rho}$ we would
obtain, in Fig. \ref{radata}, curves with the same aspect of the solid line but
shifted upwards. For conciseness we decided not to include them in the figure. 
Furthermore, the used numerical values could be different, such as the hadronization
temperature, $T_h$, or the numbers contained in the parametrizations. None of these 
changes however would susbtancially change the curve shown in Fig. \ref{radata}.

To summarize: we have improved the treatment of the microscopic dynamics of
$K^*$'s. We used all the relevant reaction cross sections involving $K^*$'s
calculated in Ref. \cite{abreu} as input in the evolution equations
(\ref{sistema}). We included cooling and the dependence of the freeze-out
temperature on the system size. We obtained a very good description of the data
published in \cite{alice20} on $R = K^*/K$ as a function of
$d N / d \eta (\eta=0)$. In order to reproduce the features of Fig. \ref{radata}
we need the three aspects of the process: i) dominance of the $K^*$ absorption
reactions; ii) cooling and iii) system size dependent freeze-out.

\section{Appendix}

In this appendix we have included the parametrization used to reproduce the
thermally averaged cross sections calculated in Ref. \cite{suhoung} and in
Ref. \cite{abreu}. It is given by:
\begin{equation}
\braket{\sigma \,  v} (T) =   p_0 + p_1 T + p_2 T^2 + p_3 T^3 .
\label{param}
\end{equation}
The coefficients $p_i$ are given in Table I. 
\begin{table}[!ht] 
\begin{tabular}{ccccc}
  \hline
& $p_0$         & $p_1$       & $p_2$      & $p_3$      
\\ \hline \hline
$K^* \rho \to K \pi $ \cite{abreu}      & 92     & -0.91    &  0.0043
& $-7.2 \times 10^{-6}$                     \\ \hline
$K^* \rho \to K \pi $ \cite{suhoung}    & 1.78   & -0.0052  & 0.000007
& 0                                         \\ \hline
$K \pi \to K^* \rho $ \cite{abreu}      & -20    & 0.6      & -0.007  
& $3.5 \times 10^{-5}$                      \\ \hline
$K \pi \to K^* \rho $ \cite{suhoung}    & 0.2    & -0.004   & 0.00001
& $6.0 \times 10^{-8}$                      \\ \hline
$K^* \pi \to K \rho $ \cite{abreu}      & -8.5   & 0.200    & -0.00085 
&  $1.23 \times 10^{-6}$                    \\ \hline
$K^* \pi \to K \rho $ \cite{suhoung}    & -0.1   & -0.002   & 0.00007
&  $-1.5 \times 10^{-7}$                    \\ \hline
$K \rho \to K^* \pi $ \cite{abreu}      & 25.3   & -0.143   & 0.00052  
&  $-8.0 \times 10^{-7}$                    \\ \hline 
$K \rho \to K^* \pi $ \cite{suhoung}    & 0      & 0.010    & -0.000014
& 0                                         \\ \hline
$ K \pi \to K^*     $ \cite{suhoung}    & -3     & 0.27     & -0.0019
& $3.8 \times 10^{-6}$                      \\ \hline
$K^* \to K \pi $ \cite{suhoung}         & 0.2579 & $-4.32 \times 10^{-4}$
& $6.0 \times 10^{-7}$ & $ -6.5 \times 10^{-10}$          \\ \hline
\end{tabular}
\caption{Parameters used in (\ref{param}). With the above numbers the temperature 
  is given in MeV and the outcome are the thermally averaged cross sections in mb.
In the last line, the average decay width is given in $fm^{-1}$.}
\label{paracross}
\end{table}

\begin{acknowledgments}
This work was  partially financed by the Brazilian funding agencies CAPES
and CNPq.
\end{acknowledgments}


\end{document}